# Communicability in complex networks

Ernesto Estrada[1] & Naomichi Hatano[2]

[1]*Complex Systems Research Group, X-rays Unit, RIAIDT, Edificio CACTUS, University of Santiago de Compostela, 15076 Santiago de Compostela, Spain*

[2]*Institute of Industrial Science, University of Tokyo, Komaba 4-6-1, Meguro, Tokyo 153-8505, Japan*

Many topological and dynamical properties of complex networks are defined by assuming that most of the transport on the network flows along the shortest paths. However, there are different scenarios in which non-shortest paths are used to reach the network destination. Thus the consideration of the shortest paths only does not account for the global communicability of a complex network. Here we propose a new measure of the communicability of a complex network, which is a broad generalization of the concept of the shortest path. According to the new measure, most of real-world networks display the largest communicability between the most connected (popular) nodes of the network (assortative communicability). There are also several networks with the disassortative communicability, where the most "popular" nodes communicate very poorly to each other. Using this information we classify a diverse set of real-world complex systems into a small number of universality classes based on their structure-dynamic correlation. In addition, the new communicability measure is able to distinguish finer structures of networks, such as communities into which a network is divided. A community is unambiguously defined here as a set of nodes displaying larger communicability among them than to the rest of nodes in the network.

Complex networks represent interactions between pairs of units in disparate physical, biological, technological and social systems (1-4). A focus of research in this field is the search of good measures, global or local, that quantify unique characteristics of the networks (5-7). Most of the measures currently in use are based on the shortest paths connecting two units (nodes) of a network (5-7). Their relevance rests on the premise that communication between the nodes takes place through the shortest paths (8-10).

At a local scale, the shortest path is often used to identify network communities (11, 12) or to characterise the importance of the nodes in a network (13). For instance, the boundaries of a community are commonly defined (11) on the basis of the influence of a node over the flow of information between other nodes, assuming that this flow primarily follows the shortest paths. At a global scale, the use of many concepts like the average shortest path length (14), the degree-degree correlations (15) and the degree distribution (16) emphasises the communicability through the shortest paths.

However, "information" can in fact spread along non-shortest paths (14, 17). We can think, for instance, of gossip spreading in a social network, where the information can flow back and forward several times before reaching the final destination. Consequently, concepts like "small worldness" (18), "assortativeness" (19) or "scale-freeness" (16) can miss important information on the network communicability as well as on finer structures of the network depending on it (20).

## Communicability in complex networks

We consider networks represented by simple graphs $G = (V, E)$, that is, graphs having $|V| = n$ nodes and $|E| = m$ links, without self-loops or multiple links between nodes. Let $A(G) = A$ be the adjacency matrix of the graph whose elements $A_{ij}$ are ones or zeroes if the corresponding nodes $i$ and $j$ are adjacent or not, respectively. We will call the eigenvalues of the adjacency matrix in the non-increasing order $\lambda_1 \geq \lambda_2 \geq \cdots \geq \lambda_n$, the spectrum of the graph (21).

It is well-known that the $(p, q)$-entry of the $k$ th power of the adjacency matrix, $\left(A^k\right)_{pq}$, gives the number of walks of length $k$ starting at the node $p$ and ending at the node $q$ (21). A walk of length $k$ is a sequence of (not necessarily different) vertices $v_0, v_1, \cdots, v_{k-1}, v_k$ such that for each $i = 1, 2 \cdots, k$ there is a link from $v_{i-1}$ to $v_i$. Consequently, these walks communicating two nodes in the network can revisit nodes and links several times along the way, which is sometimes called "backtracking walks". In contrast, a path is a sequence of different vertices.

The *communicability* between a pair of nodes in a network is usually considered as the shortest path connecting both nodes. We now propose a generalization of the communicability by accounting not only for the shortest paths communicating nodes $p$ and $q$ but also for all the other walks that permit for a "particle" to travel from one to the other.

The theoretical justification for this consideration is two-fold. First, it is known that communication between a pair of nodes in a network does not always take place through shortest paths but it can follow non-shortest paths routes. The other justification is that the shortest paths are not very sensitive with respect to the appearance of structural bottlenecks in a network. On the contrary, the number of walks is significantly affected by the appearance of these structural changes in a network.

Our strategy here is to make longer walks have lower contributions to the communicability function than shorter ones. If $P_{pq}^{(s)}$ is the number of the shortest paths between the nodes $p$ and $q$ having length $s$ and $W_{pq}^{(k)}$ is the number of walks connecting $p$ and $q$ of length $k > s$, we propose to consider the quantity

$$G_{pq} = \frac{1}{s!}P_{pq} + \sum_{k>s}\frac{1}{k!}W_{pq}^{(k)}.$$ **[1]**

While a shortest path represents only a single path that communicates both nodes, our approach considers all ways in which we can reach the target node $q$ starting our walk at the node $p$. As some of these "detours" can be very long, the summation is weighted in decreasing order of the length of the walk.

Using the connection between the powers of the adjacency matrix and the number of walks in the network, we obtain

$$G_{pq} = \sum_{k=0}^{\infty}\frac{\left(A^k\right)_{pq}}{k!} = e^A.$$ **[2]**

This can be further rewritten in terms of the graph spectrum as (22)

$$G_{pq} = \sum_{j=1}^{n}\varphi_j(p)\varphi_j(q)e^{\lambda_j},$$ **[3]**

where $\varphi_j(p)$ is the $p$ th element of the $j$ th orthonormal eigenvector of the adjacency matrix associated with the eigenvalue $\lambda_j$ (21). We will call $G_{pq}$ the *communicability* between the nodes $p$ and $q$ in the network.

## Communicability as the Green's function of networks

We now argue that the communicability defined above is actually the Green's function of the network. For a given network with the adjacency matrix $A$, imagine the following system. We have a spring on each link of the network. We somehow put the network of springs on a plane, adjusting the natural length of the springs so that the system may be at rest on the plane. Each node can oscillate in the direction perpendicular to the plane. The $p$th node, when at height $z_p$, feels the force $F_p = K \sum_q (z_p - z_q) A_{pq}$, where $K$ is the common spring constant, because the $p$th node is connected by a spring to the $q$th node only if $A_{pq} = 1$. In other words, the potential energy for the $p$th node is given by

$$U_p = \frac{K}{2} \sum_q (z_p - z_q)^2 A_{pq} \qquad [4]$$

and hence the total energy is given by

$$E = \sum_p U_p = \frac{K}{2} \sum_{p,q} (z_p - z_q)^2 A_{pq}, \qquad [5]$$

which after some algebraic manipulation is transformed in the expression

$$E = K \left( \sum_p z_p^2 k_p - \sum_{p,q} z_p A_{pq} z_q \right) = -K \sum_{p,q} z_p L_{pq} z_q, \qquad [6]$$

where $k_p$ is the degree of the $p$th node, or the number of links attached to the $p$th node, and $L_{pq}$ is the corresponding element of the Laplacian matrix of the graph $L_{pq} = A_{pq} - k_p \delta_{pq}$. The partition function is given by

$$Z = \sum_{\text{all configurations}} e^{-\beta E} = \int \exp\left( \beta K \sum_{p,q} z_p L_{pq} z_q \right) \prod_q dz_q. \qquad [7]$$

We can transform the partition function in terms of the normal modes. Suppose that we diagonalize the Laplacian matrix $\boldsymbol{L}$ in the form $\sum_q L_{pq} \varphi_j^{(q)} = \lambda_j \varphi_j^{(p)}$. Then the partition function **7** is transformed to

$$Z = \int \exp\left( \beta K \sum_j \lambda_j u_j^2 \right) \prod_j du_j, \qquad [8]$$

where $u_j = \sum_q z_q \varphi_j^{(q)}$. The integration in **8** is now possible, being the product of Gaussian integrals. Let us now calculate the correlation function, or the (thermal) Green's function

$$G_{pq}(\beta) = \langle z_p z_q \rangle = \frac{1}{Z} \int z_p z_q \exp\left( \beta K \sum_{s,t} z_s L_{st} z_t \right) \prod_r dz_r. \qquad [9]$$

After the same transformation above, we obtain

$$G_{pq}(\beta) = \langle z_p z_q \rangle = \sum_j \varphi_j^{(p)} \varphi_j^{(q)} e^{\beta K \lambda_j} = \sum_{k \geq s}^{\infty} \frac{\beta^k}{k!} W_{pq}^{(k)}. \qquad [10]$$

This describes how much the $q$th node oscillates when we shake the $p$th node.

In general, the Green's function expresses how an impact propagates from one place to another place. In this sense, Eq. **10** is nothing but the Green's function of the network. From another point of view, we can consider particle diffusion on the complex network. Then the Green's function **10** describes how many particles end up at the $q$th node if we put particles at the $p$th node.

Hereafter, we ignore the diagonal elements of the matrix $\boldsymbol{L}$ and simply used the adjacency matrix $\boldsymbol{A}$ for the total energy. This corresponds to introducing additional springs that connect each node to the two-dimensional plane itself so that we can cancel the inhomogeniety of the diagonal elements. We also restrict ourselves to the case $\beta = 1$ in what follows.

## Degree-communicability correlations

In order to investigate the structure-dynamic relationship in complex networks, we use the correlation between the node degree and the communicability (the Green's function). The node degree $k_p$ is one of the simplest topological characteristics of a network defined as the number of links attached to a node. The correlation can be observed in the form of three-dimensional contours where $k_p$ and $k_q$ form the $x$ and $y$ axes, and $G_{pq}$ is plotted as the $z$. We then fit the data points by using the weighted least square method, which is implemented in the STATISTICA package. This method is similar to the one proposed by McLain for drawing contours from arbitrary data points (23).

According to the degree-communicability pattern networks can be classified in any of the following three classes:

Class (a): Homogeneous networks with assortative communicability;

Class (b): Non-homogeneous networks with assortative communicability;

Class (c): Non-homogeneous networks with disassortative communicability.

Assortative communicability (AC) is the characteristic of a network of communicating according to an assortative pattern, in which the largest communicability takes place among the nodes with the highest degrees (hubs) and the lowest communicability occurs between nodes of low degree. On the other hand, disassortative communicability (DC) is the pattern in which the largest communicability occurs between hubs and nodes of low degree. In DC the communicability between hubs is very poor as well as among nodes of low degree.

We first consider three typical network structures, which are shown in Fig. 1 together with their $\left(k_p, k_q, G_{pq}\right)$-plots for every pair of nodes $(p, q)$. The node degree $k_p$ denotes the number of the links attached to the node $p$.

**Insert Fig. 1 here.**

The contour plot in Fig. 1*a* fits intuitive interpretation; the communicability $G_{pq}$ is high between pairs of hubs, or nodes of high degree. This pattern of communicability will be designated as the assortative communicability (AC) hereafter, because the nodes communicate preferentially to other nodes with similar connectivity characteristics. AC can appear in very homogeneous networks where the hubs can communicate to each other without structural bottlenecks (see Fig. 1*a*).

In some situations, networks with bottlenecks can also display AC. Two typical examples are a network where most of the hubs are located in one of the tightly connected clusters and a network where the hubs are the bottlenecks (Fig. 1*b*).

The contour plot in Fig. 1*c* might be counterintuitive. In social networks terminology (13), it is equivalent to saying that the most popular people are poorly communicated among them. This situation emerges when there are a couple of leaders, each of whom forms a community of many followers. The communication between the communities can be bad, and hence there is poor communicability between the leaders.

Among the 50 real-world networks that we studied (see Supplementary Information), we found 38%, 50% and 12% in each of three classes represented in Fig. 1, respectively. In Fig. 2 we show contour plots for some of these real-world networks: (a) the airport network in the USA; (b) the semantic network of the Roget's thesaurus; (c) the food web of Bridge Brook; (d) the direct transcription network between genes of yeast (*S. cereviciae*); (e) the social network of injecting drug users (IDUs); (f) the social network of people with HIV infection in Colorado Spring during the period of 1985-1999.

**Insert Fig. 2 here.**

The first two networks (Fig. 2*a* and *b*) clearly display AC. The USA airport network is characterized by the lack of topological bottlenecks (24). This structural homogeneity results in the high inter-hub communicability of Class (a) as in Fig. 1*a*. The Roget's thesaurus network also displays AC despite it is formed by several clusters separated by structural bottlenecks (24). In this case, however, there is a preference of the hubs to be connected to other hubs, and hence we have Class (b) as in Fig. 1*b*.

The food web in Fig. 2*c* forms a homogeneous network without large structural bottlenecks (25). However, this network shows very large preference of the hubs to be attached to low degree nodes. Consequently, most of the inter-hub communication takes place by indirect routes decreasing the inter-hub communicability.

The last three cases, Fig. 2*d—f*, display some degrees of DC of Class (c) as in Fig. 1*c*; the largest communicability takes place between a hub and a node of low degree. They are highly clustered networks (24), but this characteristic alone is not able to explain their DC patterns. Networks such as the protein-protein interaction network of yeast and the transcription network of *E. coli* are also highly clustered (4) but display AC characteristics; they have different clusters but the hubs in each of them are directly connected to each other as in Fig. 1*b*. Then, how can we have the DC patterns? The network of injecting drug users (Fig. 2*e*) has a core of tightly connected individuals that interchange needles among them. This core is formed by several hubs, i.e., individuals that share their needles with a large number of other users. These hubs interchange their needles among them giving rise to certain AC characteristics observed in the contour plot of Fig. 2*e*. However, there are several other groups in the network lead by other individuals with large number of internal connections. These groups are almost isolated and communicate among them only through very few individuals. This gives rise to the DC characteristics observed in Fig. 2*e*. In the case of the risk network of Colorado Spring there is not a highly interconnected core (26) and the network shows very clear DC characteristics.

## Method of identifying network communities

We now present a method of analyzing the structure of a complex network. More specifically, we show how we can identify network communities by using the communicability, or the Green's function. Community identification has been an active area of research in complex networks (11, 12, 27-31).

In order to make further analysis, we now use the spectral decomposition of the Green's function (3). Imagine again that the network has a spring on its each link. Each eigenvector indicates a mode of oscillation of the entire network and its eigenvalue represents the weight of the mode. It is known that the eigenvector of the largest eigenvalue $\lambda_1$ has elements of the same sign. This means that the most important mode is the oscillation where all nodes move in the same direction at one time.

The second largest eigenvector $\varphi_2$ has both positive and negative elements. Suppose that a network has two clusters connected through a bottleneck but each cluster is closely connected within. The second eigenvector represents the mode of oscillation where the nodes of one cluster move coherently in one direction and the nodes of the other cluster move coherently in the opposite direction. Then the sign of the product $\varphi_2(p)\varphi_2(q)$ tells us whether the nodes $p$ and $q$ are in the same cluster or not.

The same analysis can be applied to the rest of the eigenvalues of the network. The third eigenvector $\varphi_3$, which is orthonormal to the first two eigenvectors, have a different pattern of signs, dividing the network into three different blocks after appropriate arrangement of the nodes. The second eigenvector divides the graph into biants, the third divides it into triants, the fourth into quadrants, and so forth.

According to this pattern of signs we have the following decomposition of the thermal Green's function:

$$G_{pq} = \varphi_1(p)\varphi_1(q)e^{\lambda_1} + \sum_{j=2}^{n} \varphi_j^+(p)\varphi_j^+(q)e^{\lambda_j} + \sum_{j=2}^{n} \varphi_j^-(p)\varphi_j^-(q)e^{\lambda_j} + \sum_{j=2}^{n} \varphi_j^+(p)\varphi_j^-(q)e^{\lambda_j} \text{ [11]}$$

where $\varphi_j^+$ and $\varphi_j^-$ refer to the eigenvector components with positive and negative signs, respectively. The first three terms on the right-hand side of **11** give positive contributions and the last term makes a negative contribution to the thermal Green's function. According to the partitions made by the pattern of signs of the eigenvectors in a graph, two nodes have the same sign in an eigenvector if they can be considered as being in the same partition of the network, while those pairs having different signs correspond to nodes which are in different partitions. Thus, the second and third terms of **11** represent the *intra-cluster communicability* between nodes in the network and the last term represents the *inter-cluster communicability* between nodes.

The above consideration motivates us to define a quantity $\Delta G_{pq}$ by subtracting the contribution of the largest eigenvalue $\lambda_1$ from Eq. **2**, or removing the background mode of translational movement. Then the positive contributions to the sum in $\Delta G_{pq}$, indicating that the nodes $p$ and $q$ are in the same cluster, represent the intra-cluster communicability. The negative contributions, on the other hand, indicate that the nodes $p$ and $q$ are in different clusters, and hence represent the inter-cluster communicability:

$$\Delta G_{pq}(T) = \sum_{j=2}^{\text{intra-cluster}} \varphi_j(p)\varphi_j(q)e^{\beta\lambda_j} + \sum_{j=2}^{\text{inter-cluster}} \varphi_j(p)\varphi_j(q)e^{\beta\lambda_j}. \qquad \text{[12]}$$

By focusing on the sign of $\Delta G_{pq}$, we can unambiguously define a community as a group of nodes. If $\Delta G_{pq}$ for a pair of nodes $p$ and $q$ have a positive sign, they are in the same community. If $\Delta G_{pq}$ for the two nodes have a negative sign they are in different clusters.

As we are considering every pair of nodes in the network we can represent the network as a signed complete graph. A signed complete graph is a graph in which every pair of nodes are linked to each other and every link in the graph has a positive or negative sign. Thus, it is straightforward to realize that a community is a positive clique in the signed complete graph. A positive clique is a subgraph in which every pair of nodes are linked to each other and all links have a positive sign. Then, a community can be formally defined as the largest possible positive clique in the signed complete graph. Consequently, the method of detecting networks communities is reduced to find these maximal positive cliques.

Figure 3*b* is the signed complete graph for the network in Fig. 3*a*. In Fig. 3*c* we illustrate the four positive cliques extracted from this signed graph. The maximal positive clique that can be formed by the nodes 1, 2, 3, 4 and 5 is the 5-clique, which represents a community formed by the five nodes. However, the maximal positive cliques formed by the nodes 5, 6 and 7 are a couple of 2-cliques forming the clusters 3 and 4; there is not a positive 3-clique formed by these nodes.

**Insert Fig. 3 here.**

By representing the signs of the values of $\Delta G_{pq}$ in a matrix, we obtain a signed matrix as in Fig. 3*d*. After appropriate rearrangement of the rows and columns of this matrix we see that every community is represented by a square positive sub-matrix. The communities found using this approach for the network under analysis are illustrated in Fig. 3*e*, where we can see that the current method not only identifies simple communities but also their overlapping. In addition, the values of $\Delta G_{pq}$ (not the sign) can be used as a criterion of the cohesiveness of a community. The larger the values of $\Delta G_{pq}$ the tighter the relation between the corresponding members of this community.

As an example of real-world network, we consider a friendship network known as the Zachary karate club, which has 34 members (nodes) with some friendship (links). The members of the club, after some entanglement, were eventually fractioned into two groups, one formed by the followers of the instructor and the other formed by the followers of the administrator. The average Green's functions for this network are $G = 17.52$ and $\Delta G = -0.15$, where $\cdots$ stands for the average over all pairs of nodes. No pair of nodes has $\Delta G_{pq} = 0$; most of the pairs (87%) have $-2 \le \Delta G_{pq} \le 2$, while the minimum is $\Delta G_{pq} = -20.69$.

In Fig. 4, we plot the values of $\Delta G_{pq}$ for every pair of nodes in the karate club network. As can be seen in Fig. 4 the instructor (node 1) leads a group formed by the nodes represented at the bottom left part of the plot. On the other hand, the administrator (node 34) is the leader of the other faction formed by the nodes represented at the top right part of the plot.

**Insert Fig. 4 here.**

As is suggested in Fig. 3*e*, the current approach permits the identification of the overlapping between communities of nodes pertaining to more than one group simultaneously. The real-world communities characteristically display some degree of overlapping to each other (30). In the friendship network of the Zachary karate club, we identify two large communities, one formed by the followers of the instructor (node 1) and the other formed by the followers of the administrator (node 34). The nodes forming the instructor´s faction (the red circles in Fig. 4) only form one community. That is, these individuals are tightly communicated to each other in one community lead by the instructor.

However, the followers of the administrator form a more fractioned community. Not all followers of the administrator communicate very well to each other. This gives rise to several overlapped communities among these groups of individuals. For instance, in Fig. 5 we illustrate two of these communities. The first, in yellow, is formed by all blue squared nodes except nodes 9 and 31. The other community, in blue, is formed by all nodes except nodes 25 and 26. The overlap between these two communities is represented in green. It is formed by those individuals who are simultaneously in both communities. There is still another community, not represented in Fig. 5, which is formed by all nodes except nodes 9 and 25.

**Insert Fig. 5 here.**

**Conclusions**

We have extended the concept of communicability in networks beyond the simple consideration of the shortest paths connecting nodes. The conventional definition accounts only for the shortest paths as the communicability. The definition introduced here takes longer walks into account. The number of walks is measured through the powers of the adjacency matrix of the network. We define the communicability between two nodes by giving larger weights to the shorter walks and smaller weights to the longer walks. The shortest paths connecting two nodes always make the largest contribution to the communicability, but longer walks, greater in number, also have some contributions. Our definition permits analytical calculation of the communicability from graph spectral theory as well as identification of this measure as the thermal Green's function of the network. In other words, the communicability function expresses how an impact propagates from one node to another in the network.

The use of our definition of network communicability has several unique features. We can obtain information about network structures at both global and local scales simultaneously, which has been identified as a promising route to explore complex networks (32). We have shown that this information is critical to understanding the organization and evolution of complex networks. First, we have used this measure to investigate the structure-dynamic relationship in real-world complex networks. By analyzing the degree-communicability relations we have empirically discovered the existence of three universality classes of complex networks: the homogeneous networks which always display assortative communicability (AC) and the heterogeneous networks that can display either assortative or dissasortative (DC) communicability. In AC networks the most connected nodes or hubs display the largest communicability among them following the common intuition. Less intuitive is the case of DC networks in which hubs are poorly communicated among them.

Network communicability also permits an unambiguous definition of a community in a network. A community is a set of nodes in the network displaying the largest internal communicability, that is, a group of nodes that communicate much better among them than with the rest of the nodes in the network. This definition enables analytical identification of communities in a network as has been illustrated here for the Zachary karate club. An interesting feature of this method is that it permits to find overlapping communities in the network, which is closer to the real-life situation than the definition of isolated communities.

In closing, network communicability as defined here is a promising measure for analysing topological and dynamical properties of graphs and networks. The information displayed by this graph theoretical measure is not duplicated by other existing measures and its facility of calculation will permit its application in many different areas of research using graphs and networks.

**Acknowledgement.** We thank J. A. Dunne, R. Milo, U. Alon, J. Moody, V. Batagelj, J. Davis, J. Potterat, P. M. Gleiser, D. J. Watts, C. R. Myers and C. Baysal for generously providing datasets. This work was partially supported by the "Ramón y Cajal" program, Spain.

**Figure 1:** Illustration of three different organizations of nodes in networks and their communicability patterns. (*a*) Super-homogeneous network where the "information" can flow among hubs without passing through structural bottlenecks. The contour plot represents the relative communicability between every pair of nodes as function of their degrees ($k_p, k_q$). A super-homogeneous network displays the largest communicability between the most connected nodes (blue nodes) and the lowest communicability between the nodes of low degree (red nodes), i.,e, assortative communicability. (*b*) Network formed by two (or more) clusters of highly interconnected nodes which have very few inter-cluster connections (bottleneck). In this case the hubs (blue nodes) of one cluster are directly connected to the hubs of the other. Consequently, the communicability pattern is similar to the one shown in Fig. 1*a*. (*c*) Network with two (or more) clusters in which the "information" arising at the hubs (blue nodes) of one cluster needs to travel through the bottleneck to reach the hubs (blue nodes) of the other cluster. This network displays an "atypical" disassortative communicability pattern in which hubs are better communicated with nodes of low degree and the inter-hub communicability is poor.

**Figure 2:** Communicability-degree contour plots for several real-world networks. The first two plots are typical of networks with assortative communicability (AC) and the network structures correspond to cases like the ones illustrated in Fig. 1*a* and *b*. The plot in c also corresponds to AC but due to the large preference of the hubs to be attached low degree nodes the inter-hub communicability is reduced. The last three cases correspond to typical disassortative communicability (DC) patterns. The corresponding networks have structures that match the topology illustrated in Fig. 1*c*. (*a*) the airport network in the USA in 1997. (*b*) the semantic network of the Roget's thesaurus. (*c*) the food web of Bridge Brook. (*d*) the direct transcription network between genes of yeast. (*e*) the social network of injecting drug users. (*f*) the social

network of people with HIV infection in Colorado Spring during the period of 1985-1999.

**Figure 3:** Illustration of the process of identifying communities in a simple network at the top of the figure. (*a*) A representation of the signed complete graph, where the red lines indicate negative $\Delta G_{pq}$ and the blue ones indicate positive $\Delta G_{pq}$. (*b*) The four completely positive cliques existing in the network. (*c*) Identification of the communities by grouping the positive (blue) entries of the adjacency matrix. (*d*) Illustration of the different communities in the network and their overlapping.

**Figure 4:** The community structure of the Zachary karate club network. The two factions in which the network was divided are illustrated in different colours and shapes of the nodes. The matrix plot illustrates the values of $\Delta G_{pq}$ for every pair of nodes $(p,q)$ in the network. A positive value of $\Delta G_{pq}$ (reddish colour) indicates that the pair of nodes is in the same community and a negative value of $\Delta G_{pq}$ (green colours) indicates that the pair is in different communities. The nodes are ordered according to their values of $\Delta G_{pq}$ in decreasing order.

**Figure 5:** Illustration of the overlapping between two communities formed among the followers of the administrator (node 34) in the Zachary karate club network.

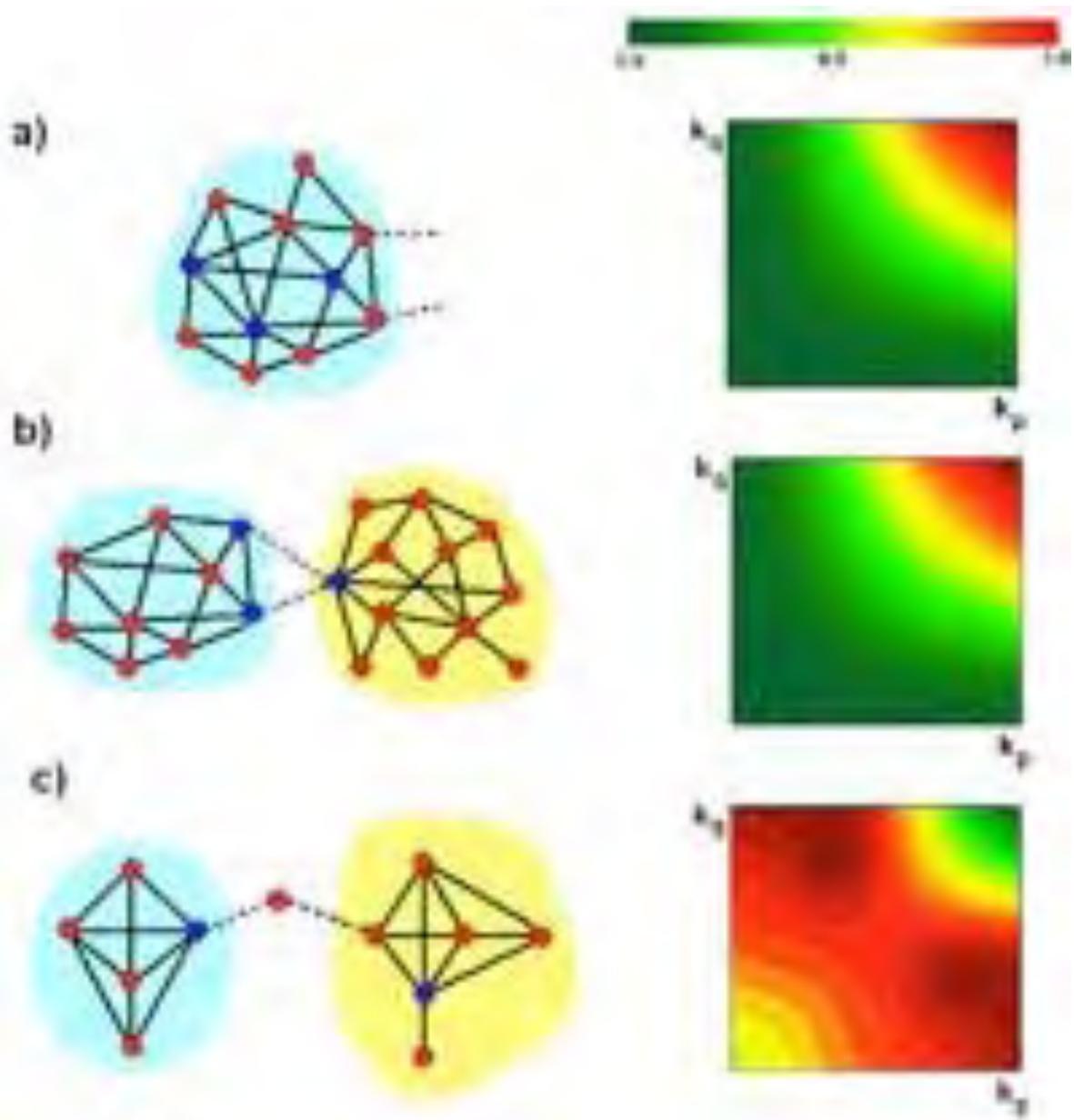

Figure 1

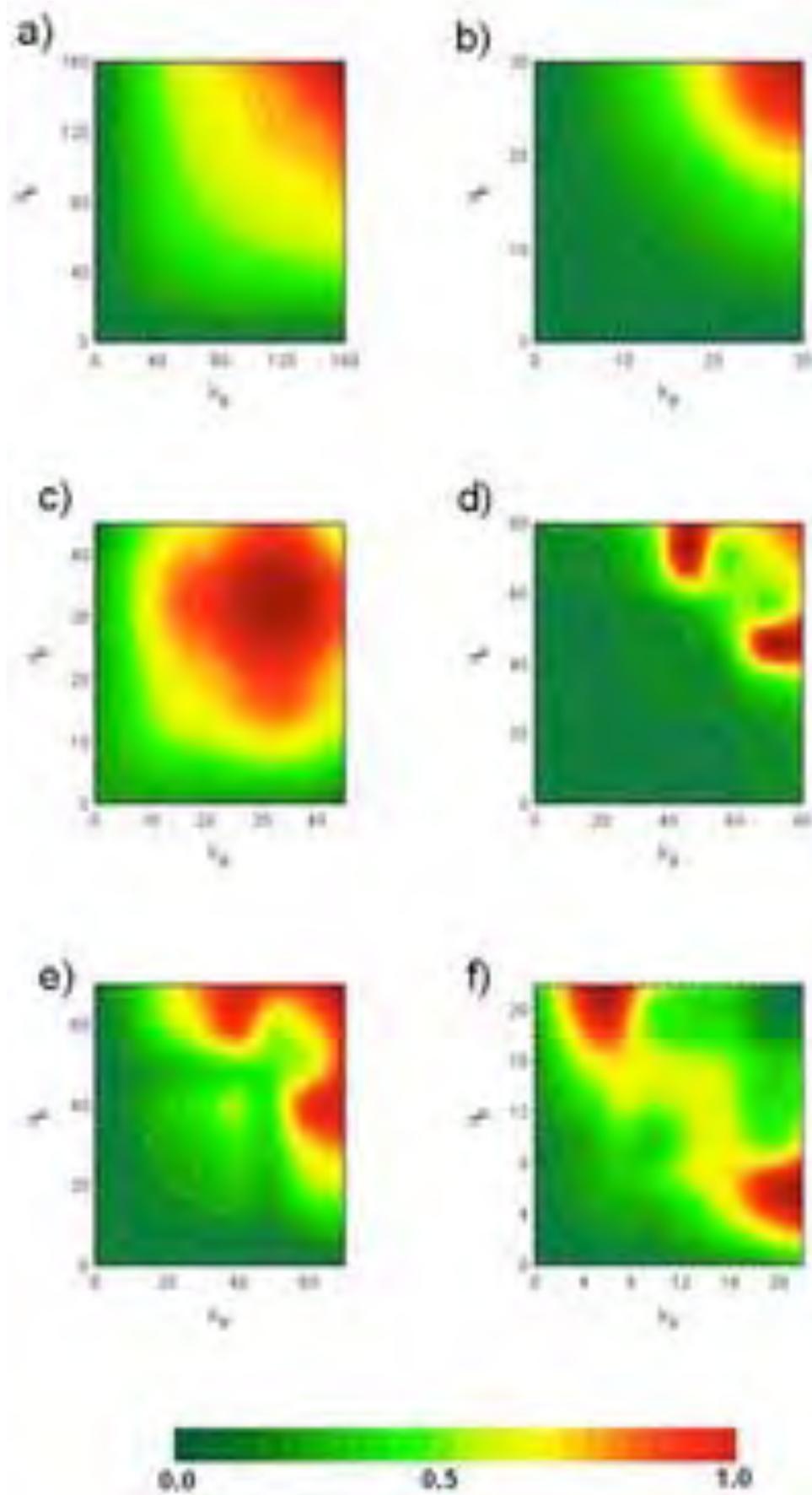

Figure 2

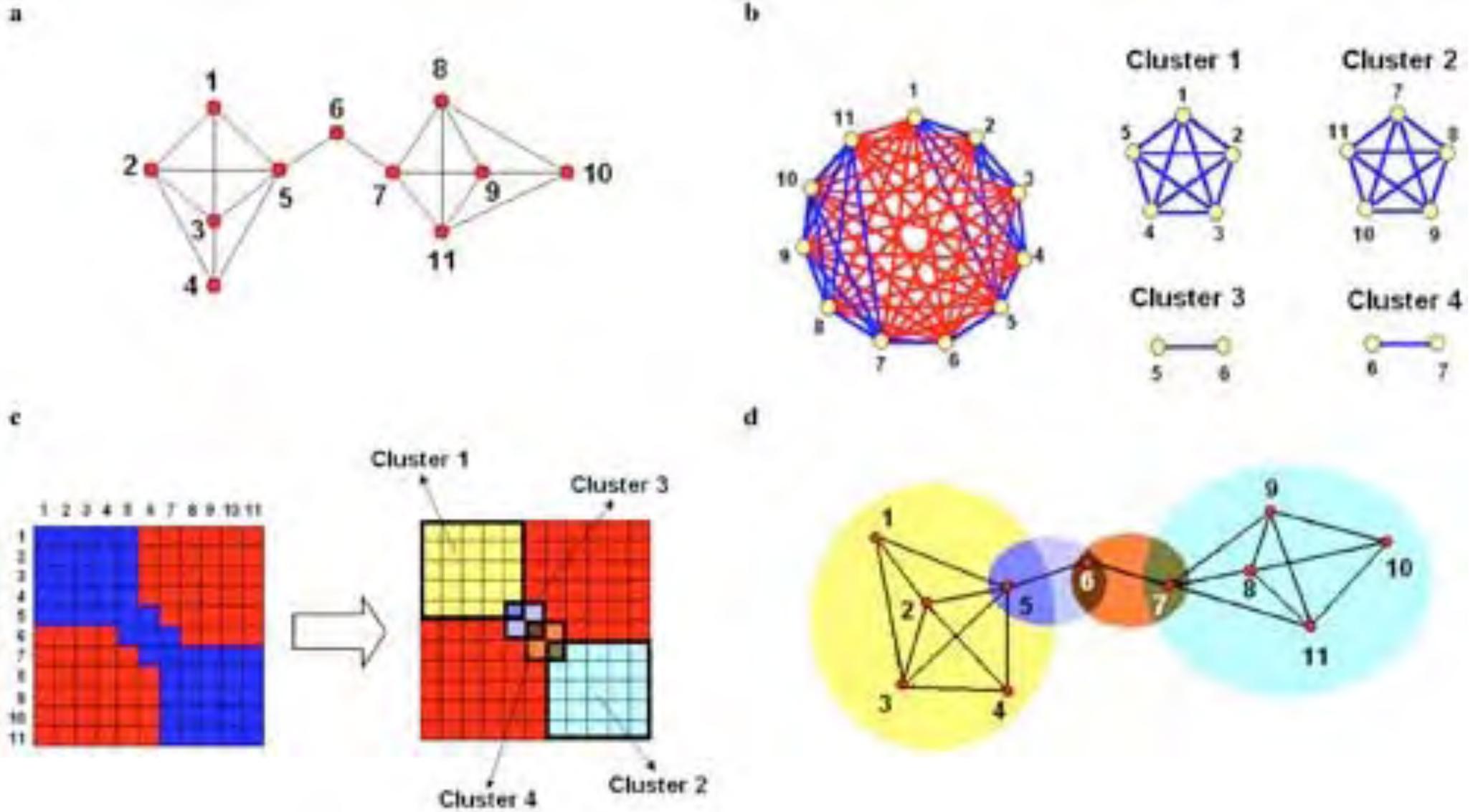

Figure 3

Figure 4

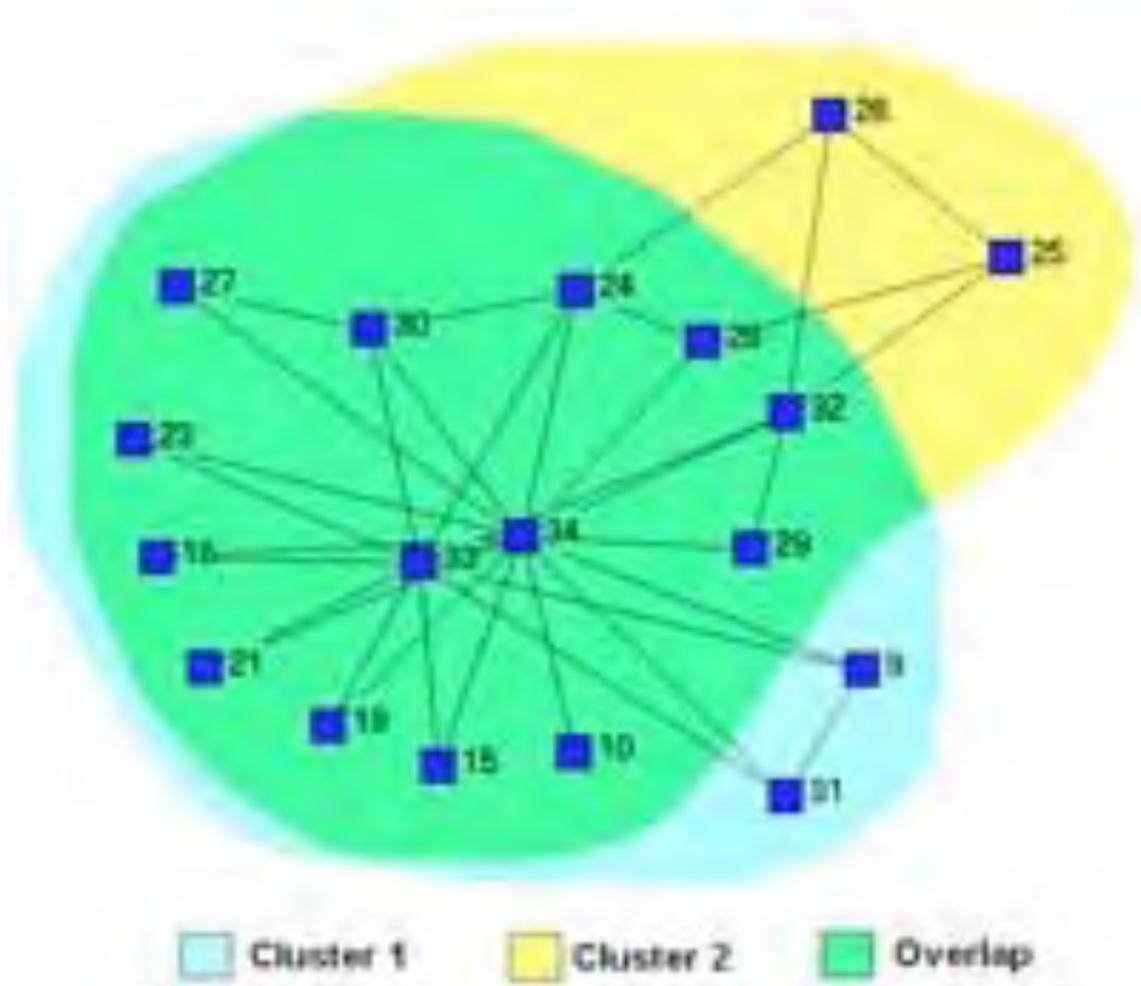

Figure 5